\pdfoutput=1
\documentclass[amsmath,prd,aps,twocolumn,showpacs,superscriptaddress,floatfix,nofootinbib]{revtex4}

\usepackage{amsfonts}
\usepackage{bm}
\usepackage{color}
\usepackage{graphicx}
\definecolor{darkgreen}{cmyk}{0.85,0.2,1.00,0.2}

\newcommand{\be}{\begin{equation}}
\newcommand{\ee}{\end{equation}}
\newcommand{\ba}{\begin{eqnarray}}
\newcommand{\ea}{\end{eqnarray}}

\newcommand\lsim{\mathrel{\rlap{\lower4pt\hbox{\hskip1pt$\sim$}}
        \raise1pt\hbox{$<$}}}
\newcommand\gsim{\mathrel{\rlap{\lower4pt\hbox{\hskip1pt$\sim$}}
        \raise1pt\hbox{$>$}}}

\def\L{{\mathcal L}}

\begin{document}

\title{On quantifying and resolving the BICEP2/Planck tension over gravitational waves}


\author{Kendrick M.~Smith}
\affiliation{Perimeter Institute for Theoretical Physics, Waterloo ON N2L 2Y5}
\author{Cora Dvorkin}
\affiliation{Institute for Advanced Study, School of Natural Sciences, Einstein Drive, Princeton, NJ 08540, USA}
\author{Latham Boyle}
\affiliation{Perimeter Institute for Theoretical Physics, Waterloo ON N2L 2Y5}
\author{Neil Turok}
\affiliation{Perimeter Institute for Theoretical Physics, Waterloo ON N2L 2Y5}
\author{Mark Halpern}
\affiliation{Dept. of Physics and Astronomy, University of British Columbia, Vancouver, BC Canada V6T 1Z1}
\author{Gary Hinshaw}
\affiliation{Dept. of Physics and Astronomy, University of British Columbia, Vancouver, BC Canada V6T 1Z1}
\author{Ben Gold}
\affiliation{Hamline University, Dept of Physics, 1536 Hewitt Avenue, Saint Paul, MN 55104}

\date{\today}


\begin{abstract}
The recent BICEP2 measurement of primordial gravity waves ($r = 0.2^{+0.07}_{-0.05}$) appears to be
in tension with the upper limit from WMAP ($r<0.13$ at 95\% CL) and Planck ($r<0.11$ at 95\% CL).  We carefully quantify the level of tension
and show that it is very significant (around 0.1\% unlikely) when the observed deficit of large-scale temperature
power is taken into account.  We show that measurements of TE and EE power spectra in the near
future will discriminate between the hypotheses that this tension is either a statistical fluke, or a
sign of new physics.
We also discuss extensions of the standard cosmological model that relieve the tension, and some novel ways to constrain them.
\end{abstract}


\maketitle


The BICEP2 collaboration's potential detection of B-mode polarization in the cosmic background radiation (CMB) has justifiably ignited enormous excitement, signaling as it may the opening of a powerful new window onto the earliest moments of the big bang~\cite{Ade:2014xna}. The implications are profound, including a possible confirmation of cosmic inflation and exclusion of rival explanations for the origin and structure of the cosmos. 

As the BICEP2 collaboration was careful to emphasize, there is some tension between their value of the parameter $r$ which 
controls the amplitude of the gravitational wave signal, 
relative to
other experiments. BICEP2 detected B-mode polarization corresponding to $r=0.2^{+0.07}_{-0.05}$ (or $r=0.16^{+0.06}_{-0.05}$ after foreground subtraction), 
as compared to upper bounds from the large-scale CMB temperature power spectrum:   $r < 0.13$ (WMAP) or $r < 0.11$ (Planck) at 95\% CL~\cite{Hinshaw:2012aka,Ade:2013zuv}.  
It is the purpose of this Letter to quantify this discrepancy in a simple manner,
to point out that measurements of CMB polarization E-modes will either sharpen or resolve it in the near future,
and to explore cosmological interpretations.

\begin{figure}
\centerline{\includegraphics[width=10cm]{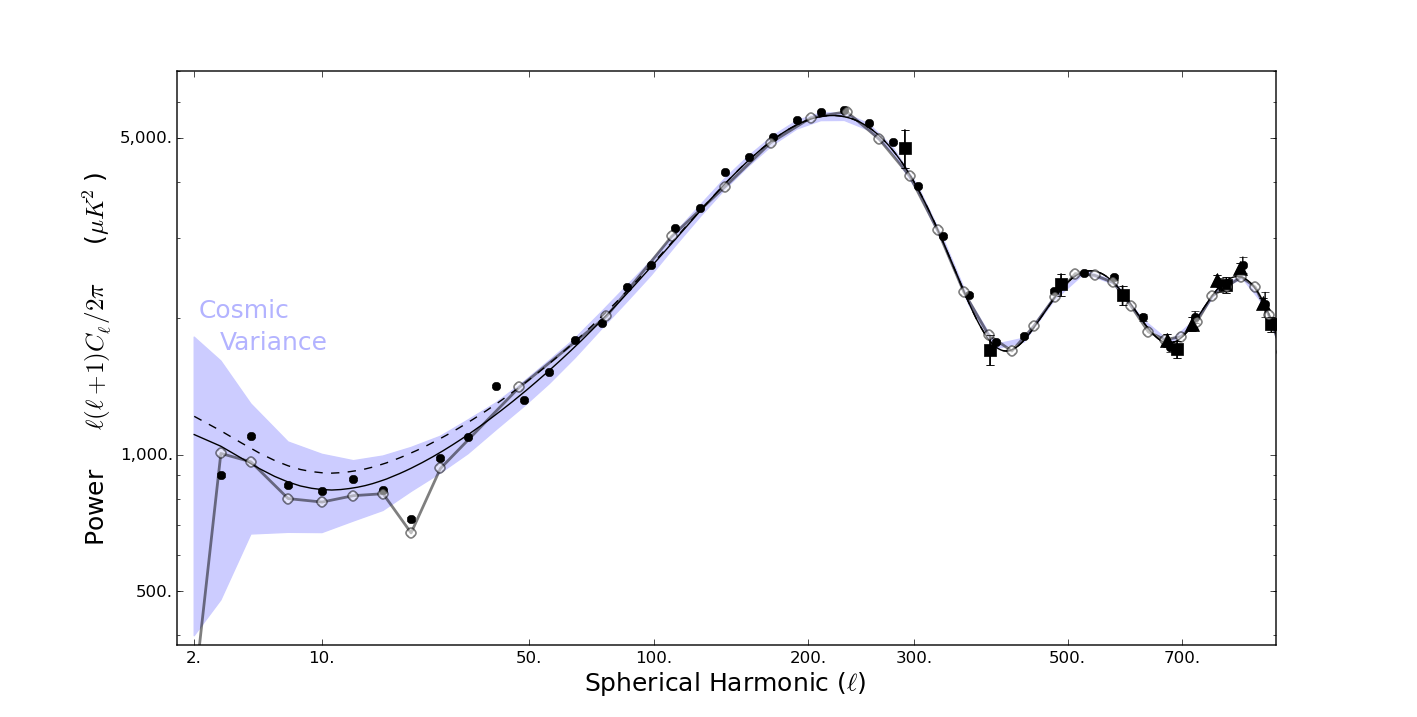}}
\caption{Current measurements of the CMB temperature power spectrum,
from Planck (open circles), WMAP (closed circles), ACT (squares) and SPT (triangles).
Error bars include noise variance only; the shaded region represents cosmic variance.
There is a small deficit of power on large angular scales relative to an $r=0$ model (solid curve)
which becomes more statistically significant if $r=0.2$ as BICEP2 suggests (dashed curve).}
\label{fig:tt}
\end{figure}

In Fig.~\ref{fig:tt}, we show current measurements of the temperature power spectrum  $C^{TT}_{l}$,
illustrating a deficit of power at low $\ell$.
This deficit was
highlighted as an important anomaly by the Planck team \cite{Ade:2013kta}. However, taken alone, it is still compatible (at the 1\% level) 
with cosmic variance and thus may be explained as a statistical fluctuation due to our only having access to a limited sample of the universe. 
BICEP2's detection of B-mode polarization, if correctly interpreted as being due to primordial gravitational waves, 
implies an additional contribution to the large-scale temperature anisotropies. 
This makes it harder to explain away the observed deficit as a statistical fluke.  

\begin{figure}[t]
\centerline{\includegraphics[width=7cm]{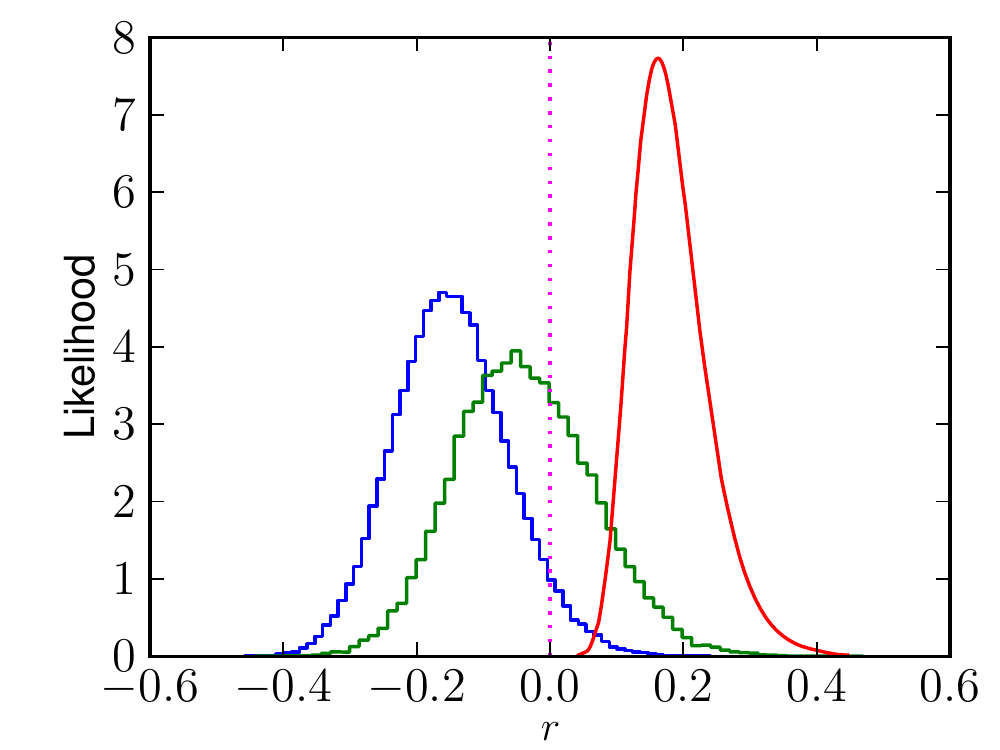}}
\caption{1D probability distribution functions for the tensor-to-scalar ratio $r$ using Planck+WP data (blue/left), WMAP+SPT+BAO+$H_0$ data (green/middle), and BICEP2 data (red/right).
We use the CosmoMC~\cite{Lewis:2002ah} code with the six cosmological parameters
\{$\Omega_b h^2$, $\Omega_m h^2$, $\Omega_\Lambda$, $A_\zeta$, $\tau$, $n_s$\}  marginalized. 
As discussed in the text, we allow $r$ to be negative in order to parametrize a possible power deficit 
on large angular scales.}
\label{fig:r_likelihoods}
\end{figure}

We quantify this problem as follows.  
We compute likelihood functions ${\mathcal L}(r)$ for $r$ 
inferred from WMAP, Planck, and BICEP2  (Fig. \ref{fig:r_likelihoods}).
Throughout this Letter, we use ``WMAP'' as a shorthand for the combination of datasets WMAP+SPT+BAO+$H_0$, and ``Planck'' as a shorthand for Planck+(WMAP polarization).
Notice that the Planck likelihood peaks at negative $r$.  Of course, $r<0$ does not make sense physically, but negative values of $r$ may be taken to provide a reasonable
parametrization of a possible deficit in low $\ell$ power, which avoids {\em a posteriori} choices in the weighting in $\ell$. 

We find that the Planck $r$-likelihood peaks $1.6\sigma$ below zero, indicating a
deficit of large-scale power.  The power deficit has been extensively studied by the
Planck collaboration~\cite{Ade:2013kta,Ade:2013zuv}; its formal statistical significance 
can be as high as 3$\sigma$ if an {\em a posteriori} choice of $\ell$-range is made.
Note that the preference for negative $r$ is hidden when an $r \ge 0$ prior is imposed 
throughout the analysis (as is typically done when quoting upper limits on $r$ from WMAP/Planck).
Indeed, a primary purpose of this Letter is to point out that the tension between Planck and BICEP2 is larger than
would be expected by comparing the $r$ constraints with an $r \ge 0$ prior imposed.

\begin{table}
\begin{center}
\begin{tabular}{|l|cc|cc|}
\hline
  &  \multicolumn{2}{c|}{$r \ge 0$ assumed}       &    \multicolumn{2}{c|}{$r < 0$ allowed}  \\
  &  WMAP & Planck & WMAP & Planck  \\  \hline\hline
No cleaning  &  0.048  &  0.007  &  0.017  &  $< 0.001$  \\
BSS cross  &  0.054  &  0.009  &  0.019  &  $< 0.001$  \\
BSS auto  &  0.067  &  0.012  &  0.024  &  $< 0.001$  \\
DDM1 cross  &  0.054  &  0.009  &  0.020  &  $< 0.001$  \\
DDM1 auto  &  0.095  &  0.020  &  0.034  &  0.001  \\
DDM2 cross  &  0.089  &  0.018  &  0.032  &  $< 0.001$  \\
DDM2 auto  &  0.189  &  0.057  &  0.066  &  0.003  \\
FDS cross  &  0.040  &  0.006  &  0.015  &  $< 0.001$  \\
FDS auto  &  0.059  &  0.010  &  0.021  &  $< 0.001$  \\
LSA cross  &  0.052  &  0.008  &  0.019  &  $< 0.001$  \\
LSA auto  &  0.059  &  0.010  &  0.021  &  $< 0.001$  \\
PSM cross  &  0.046  &  0.007  &  0.017  &  $< 0.001$  \\
PSM auto  &  0.114  &  0.026  &  0.041  &  0.001  \\
\hline
\end{tabular}
\end{center}
\caption{Probability measure of the tension between Planck/WMAP and BICEP2 results, computed from the $r$
likelihoods using Eq.~(\ref{eq:pvalue}).
(Low probabilities indicate tension.) The probability depends on whether we use Planck or WMAP data,
whether we integrate over $r < 0$, and which of the polarized dust models described in~\cite{Ade:2014xna} is used.
As we have argued in the text, integrating over $r<0$ takes the observed deficit of large-scale power into account,
and gives $\approx 3\sigma$ tension with Planck regardless of the dust model.}
\label{tab:pte}
\end{table}

To quantify the level of tension%
, we temporarily imagine that our cosmological model contains two independent parameters $r_T$ and $r_B$, 
such that $r_T$ determines the gravitational wave contribution to $C_\ell^{TT}$ and $r_B$ determines the amplitude of $C_\ell^{BB}$. 
We obtain a likelihood $\L(r_T)$ from Planck (or WMAP) and a likelihood $\L(r_B)$ from BICEP2, as shown in Fig.~\ref{fig:r_likelihoods}.
Treating these likelihoods as independent, which is justified since $T$ and $B$ are uncorrelated, the joint likelihood in the $(r_B,r_T)$-plane is obtained by multiplying them. 
If the joint likelihood has most of its support below the diagonal $r_T=r_B$, this provides a statistically significant detection of a deficit in $r_T$ relative to $r_B$. Thus we quantify
the statistical significance of the tension by computing the probability
\be
\frac{ \int_{r_T > r_B} dr_T \, dr_B \, \L(r_T) \L(r_B) }{ \int dr_T \, dr_B \, \L(r_T) \L(r_B) }  \label{eq:pvalue}
\ee
The closer this probability is to zero, the larger the tension between Planck/WMAP and BICEP2.


The results of this analysis are shown in Tab.~\ref{tab:pte}. It is seen that if we integrate over negative values of $r$ and use Planck data, then the statistical significance of the tension 
is around 3$\sigma$. Our perspective is that integrating over negative $r$ is sensible, since the observed deficit of TT
power (relative to an $r=0$ model) should contribute to the statistical significance of the tension. 
Indeed, we will see shortly that the Planck/BICEP2 tension can be interpreted as $\approx 3\sigma$ evidence for certain extensions of the 7-parameter model: either nonzero running $\alpha = dn_s/d\log k$, a blue tensor tilt $n_t$, or a higher effective number of relativistic species, which suggests that the ``true'' tension is around 3$\sigma$.

\begin{figure}
\centerline{\includegraphics[width=7cm]{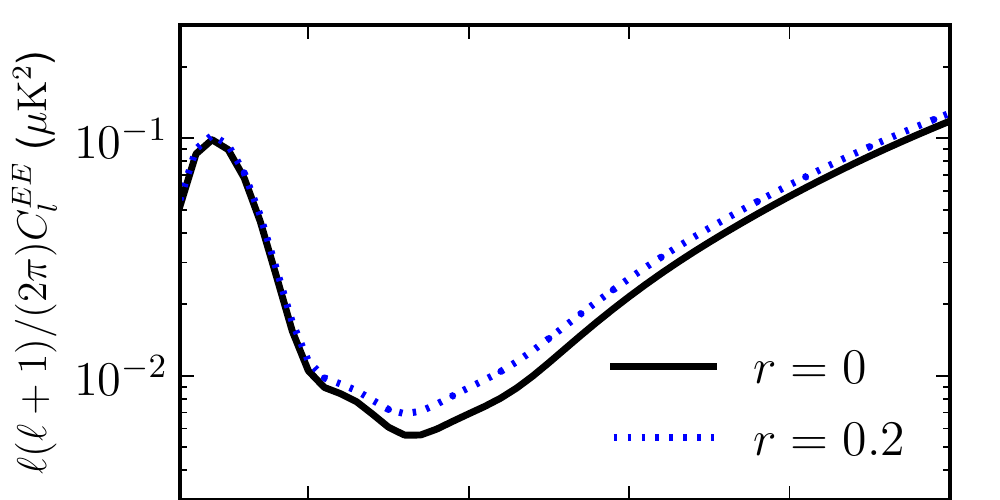}}
\centerline{\includegraphics[width=7cm]{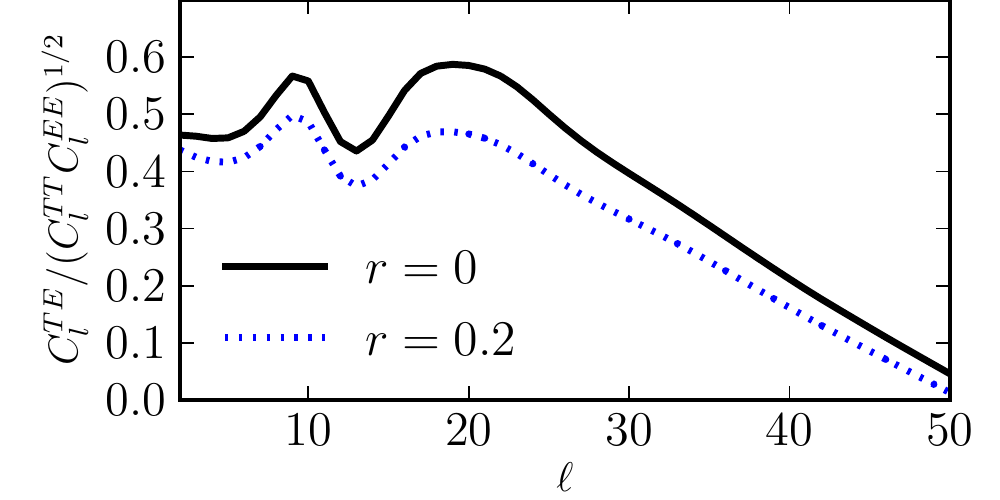}}
\caption{E-mode power spectrum $C_\ell^{EE}$ (top) and dimensionless TE correlation $C_\ell^{TE} / (C_\ell^{TT} C_\ell^{EE})^{1/2}$ (bottom) compared for $r=0$ and $r=0.2$.
An $r=0.2$ signal boosts $C_\ell^{EE}$ by $\approx 30$\% in the range $15 \lsim \ell \lsim 30$, making E-modes
more sensitive to the tensor-to-scalar ratio $r$ than temperature.}
\label{fig:emodes}
\end{figure}

While this level of tension is not so high that a definite conclusion can be drawn, it is high
enough to be worth exploring further.
Since large-scale temperature measurements are already sample variance limited, the {\it only} way to 
improve statistical errors is by measuring additional large-scale modes.  A natural source of such modes is the $E$-mode polarization of the CMB. 
The influence of $r$ on $E$-modes is particularly significant for multipoles in the range $15 \lsim \ell \lsim 30$ due to differences in the behavior 
of scalar and tensor perturbations just after horizon-crossing \cite{Coulson:1994qw, Crittenden:1994ej}.  
In this $\ell$ range,  an $r=0.2$ gravitational wave background boosts  $C_\ell^{EE}$ by $\approx 30$\% 
and suppresses  $C_\ell^{TE}$ by a similar amount (see Figure~\ref{fig:emodes}).

\begin{figure}
\centerline{\includegraphics[width=7cm]{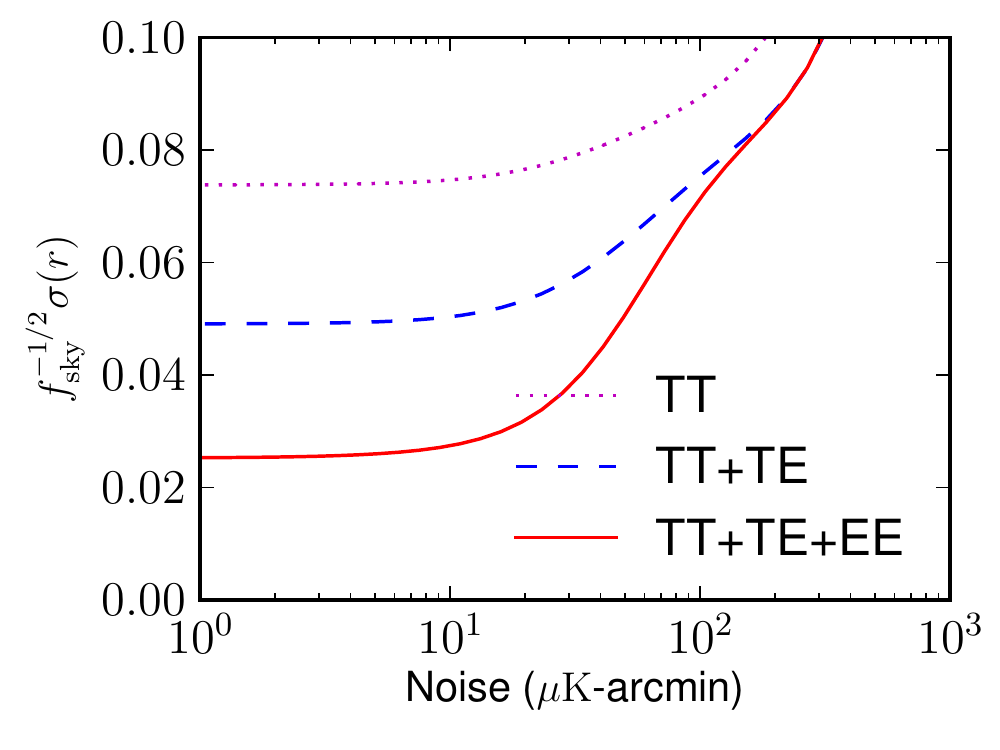}}
\caption{Statistical errors on $r$ obtained from different combinations of temperature and E-mode power spectra, as a function of noise level.
The six parameters $\{ \Omega_b h^2, \Omega_c h^2, \Omega_\Lambda, A_s, n_s, \tau \}$ have been marginalized.}
\label{fig:emodesforecast}
\end{figure}

In a scenario where the BICEP2 result holds up and and we are left wondering whether the
TT deficit is a $\approx 0.1\%$ statistical fluke or a sign of new physics, EE and TE become very interesting since they can discriminate between the two hypotheses
(a similar point was made recently by~\cite{Miranda:2014wga}).
If the TT deficit is not a statistical fluke, the deficit should be more significant when TE and EE are included.
This is illustrated in Fig.~\ref{fig:emodesforecast}, where we forecast the statistical error on $\sigma(r)$ which will result from measurements of TE and EE.
The forecasted polarization sensitivity of Planck is 70 $\mu$K-arcmin~\cite{Planck:2006aa}, and the width of the Planck+WP likelihood in Fig.~\ref{fig:r_likelihoods} is $\sigma(r) = 0.10$.
We therefore see from Fig.~\ref{fig:emodesforecast} that the statistical error on the tensor-to-scalar ratio may be reduced by a factor of two or so with Planck, and even
further with future experiments~\cite{Wu:2014hta}.
Should Planck and BICEP's central values remain unchanged, the evidence for a deficit of large-scale power as may conclusively sharpen. 
Conversely, if TE and EE measurements by Planck and other experiments provide evidence of additional large-scale power, the current tension between Planck and BICEP may be completely resolved. 
It will be fascinating to follow these developments. 

So far, we have discussed the alternative hypotheses that the Planck/BICEP2 tension is either
a statistical fluke or a symptom of needing more parameters in our cosmological model.
We conclude by briefly exploring some specific choices of extra parameters which can relieve the tension.

First we identify interesting parameters by the following approach.
For each candidate parameter
\be 
\{ \Omega_K, w, N_{\rm eff}, m_\nu, \alpha, Y_{\rm He}, n_t \}   \label{eq:candidate_parameters}
\ee
we define the parameter to be ``interesting'' if the goodness-of-fit of the Planck+WP+BICEP2
dataset to the model improves by more than $\Delta(\log {\mathcal L})=2$ when the new parameter
is included.
With this definition, we find using CosmoMC that the parameters which are interesting are the
tensor tilt $n_t$, running $\alpha = dn_s/d\log k$, and effective number of relativistic
species $N_{\rm eff}$.  We briefly explore each of these possibilities.

\begin{figure}
\centerline{\includegraphics[width=7cm,clip=true,trim=1cm 5cm 1cm 4cm]{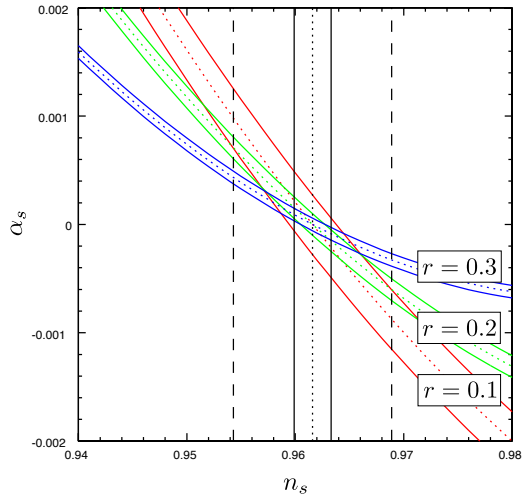}}
\caption{Forecasted 1 $\sigma$ errors in the $(n_{s},\alpha_{s})$ plane,
obtained by combining the "standard Big Bang Observer" mission proposal
\cite{BBOproposal}) with a CMB detection of $r$ of $0.1$ (red), $0.2$ (green) or
$0.3$ (blue).  The vertical lines show the current Planck value for $n_s$
(dotted), the corresponding 1$\sigma$ errors (dashed), and the forecasted
1$\sigma$ errors (solid) from a future cosmic-variance limited map of $T$ and $E$,
with $f_{{\rm max}}=0.7$ and $l_{{\rm max}}=3000$.}
\label{fig:BBOalphaFig}
\end{figure}

As pointed out by the BICEP2 collaboration, the tension with Planck may be relieved if
the running of the spectral index $\alpha_{s} = dn_s / d\log k$ is nonzero (see
also~\cite{Abazajian:2014tqa,Contaldi:2014zua,Ashoorioon:2014nta}).  The combination 
of Planck and BICEP2 data prefers negative running at almost 3$\sigma$, with a best-fit value
around $\alpha_{s}\approx-0.028$.  This is around 100 times 
larger than single-field inflation would predict, but can be realized
if $V'''/V$ is 
about 100 times larger than would 
be naively expected based on the size of $V'/V\sim (10 M_{pl})^{-1}$ and 
$V''/V\sim (10 M_{pl})^{-2}$.  Such a large value for $V'''$ makes a leading-order 
contribution to the scalar bispectrum (in contrast to the usual case, where it is 
subleading \cite{Maldacena:2002vr}); but this new contribution still 
seems too small to be detected (roughly $f_{NL} \approx 10^{-2}$).
It may be possible to measure $\alpha_s$ with statistical error approaching $10^{-3}$
in future large-scale structure experiments, if nonlinearity and bias can be modeled
to sub-percent accuracy at $k\sim 0.1$ Mpc$^{-1}$~\cite{Font-Ribera:2013rwa}.

It is clear that 
a negative $\alpha_{s}$ suppresses the small-scale scalar power.  It is perhaps
less obvious that, if the slow-roll consistency relations are satisfied, such negative 
$\alpha_{s}$ also leads to a suppression of the small-scale {\it tensor} power~\cite{consistency_paper}
that should be readily discernible by a space 
based laser-interferometric gravitational wave detector like the proposed Big Bang 
Observer (BBO) mission \cite{BBOproposal, Cutler:2009qv},
or perhaps even by the somewhat less sensitive DECIGO mission
\cite{Kawamura:2006up, Kawamura:2011zz}.  To understand this, note that we can 
use the first few slow roll consistency relations to predict the values of the tensor tilt
$n_{t}=-\tilde{r}$, the running of the tensor tilt $\alpha_{t}\equiv dn_{t}/
d\,{\rm ln}\,k=\tilde{r}(\delta n_{s}+\tilde{r})$, and the running of the running 
$\beta_{t}\equiv d\alpha_{t}/d\,{\rm ln}\,k=\tilde{r}(\alpha_{s}-\delta n_{s}^{2}
-3\tilde{r}\delta n_{s}-2\tilde{r}^{2})$, where for convenience we have defined
$\tilde{r}\equiv r/8$ and $\delta n_{s}\equiv n_{s}-1$.  
This sensitivity of $\beta_{t}$ to $\alpha_{s}$, along with 
the huge difference between $k_{{\rm BBO}}$ and $k_{{\rm CMB}}$, provides
a tremendous lever arm (roughly $k_{\rm BBO}/k_{\rm CMB} \approx 10^{17}$)
to measure $\alpha_{s}$; see Fig.~(\ref{fig:BBOalphaFig}).
If we assume the slow-roll consistency relations are satisfied,
then the CMB+BBO will be able to measure $\alpha_{s}$ with an error 
of $\pm 0.001$!
Note that the slow-roll approximation should still be satisfied on BBO
scales, since they cross the horizon at least 10 e-foldings before the
end of inflation.  The slow-roll consistency relations can be used
to extrapolate from CMB scales to BBO scales if the slow-roll parameters
obey the usual hierarchy ($\alpha_s = {\mathcal O}(\epsilon^2)$, etc.)
but fail in cases where this hierarchy is badly violated
(e.g.~models with transient features~\cite{Czerny:2014wua}).

\begin{figure}
\centerline{\includegraphics[width=7cm,clip=true,trim=0.5cm 4.5cm 1cm 3.5cm]{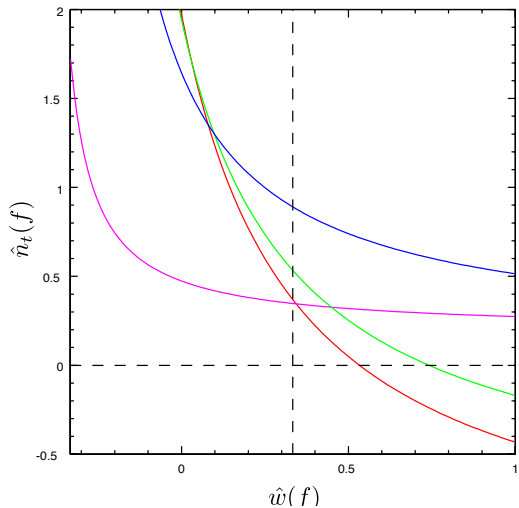}}
\caption{Upper bounds in the $\{\hat{w}(f),\hat{n}_{t}(f)\}$ plane
from combining CMB detection of $r\gtrsim0.1$ with:
current pulsar timing constraints \cite{Demorest:2012bv} (blue),
current LIGO constraints \cite{Abbott:2009ws} (green), Big Bang 
Nucleosynthesis (red), and the requirement that the primordial
tensor power is less than unity on all scales to avoid overproduction
of primordial black holes (magenta).
Here, $\hat{n}_{t}(f)$ and $\hat{w}(f)$ are appropriately averaged versions of $n_{t}(k)$ and $w(t)$:
see~\cite{Boyle:2007zx} for details.  (The standard radiation dominated scenario is $\hat{w}=1/3$.)}
\label{fig:wntFig}
\end{figure}

Next consider the tensor tilt $n_{t}$.  The Planck/BICEP2 tension can be lessened
if the tensor power spectrum is very blue ($n_t$ positive and of order one),
so that the value of $r$ measured by BICEP2 at $\ell \approx 60$ is larger than
the value of $r$ measured by Planck at $\ell \approx 30$.
The combination Planck+BICEP2 prefers positive $n_t$ at more than 3$\sigma$, with
best-fit value $n_t \approx 2.0$ (a similar result was reported in~\cite{Gerbino:2014eqa,Wang:2014kqa}).
From a theoretical standpoint, a blue tilt would be in conflict with slow roll inflation,
which predicts $n_{t}\leq0$.  From an observational standpoint, it is interesting 
to explore the extent to which a blue tilt is consistent with other constraints
on the primordial gravitational wave spectrum at much smaller comoving scales.  
Such constraints depend on the equation of state $w$ during the ``primordial dark age''
from the end of inflation to the start of BBN.
In Fig.~\ref{fig:wntFig}, we show current constraints in the $\{\hat{w}(f),\hat{n}_{t}(f)\}$
plane, where $\hat{w}(f)$ and $\hat{n}_{t}(f)$ are appropriately averaged versions
of $w$ and $n_{t}$~\cite{Boyle:2007zx}.
We see that a blue spectrum with
$n_{t}\gtrsim0.5$ runs into conflict with smaller scale constraints if it extends over too many 
decades in wavenumber. 

Finally, we have identified $N_{\rm eff}$ as an interesting parameter (as also found in~\cite{Giusarma:2014zza}).
This is best explored in combination with external datasets (ACT, SPT, $H_0$, BAO, and cluster abundance);
we refer to \cite{Dvorkin:2014lea,Zhang:2014dxk} for an in-depth discussion.

The list of candidate parameters we have considered in Eq.~(\ref{eq:candidate_parameters}) 
is not intended to be exhaustive, and it will be very interesting to consider other possibilities.
For example, a scalar field that was initially ``fast-rolling'' and then settled into its slow-roll attractor at 
around the time that the largest CMB scales left the horizon (see e.g.~\cite{Contaldi:2003zv, Cline:2003ve,Gordon:2004ez})
might produce an observationally viable scenario, with $n_{t}>0$ over a narrow range of 
scales near the horizon;
we plan to investigate this in future work.
The Planck/BICEP2 tension
is currently around 3$\sigma$, which is not yet enough to discriminate between candidate
explanations (e.g.~running, tensor tilt, unknown systematics, or a $\approx$0.1\% unlikely statistical
fluke), so at this stage we are simply enumerating possibilities.

We have carefully quantified the current tension between Planck, BICEP and the 7-parameter model, finding they are only compatible with a probability of around one in a thousand. 
In the near future, EE and TE measurements will provide a decisive test, and we will know if the tension is a statistical fluke or a sign of new physics.

\vskip 0.2cm

{\em Acknowledgements.}
We thank Wayne Hu, Stephan Meyer, David Spergel and Ned Wright for comments on the draft.
Research at Perimeter Institute is supported by the Government of Canada
through Industry Canada and by the Province of Ontario through the Ministry of Research \& Innovation.
CD was supported by the National Science Foundation grant number AST-0807444, NSF grant number PHY-088855425, and the Raymond and Beverly Sackler Funds.
LB was supported by an NSERC Discovery Grant.

\bibliographystyle{h-physrev}
\bibliography{r02tension}

\end{document}